\documentclass[aps,prl,superscriptaddress,showpacs,twocolumn]{revtex4}
\usepackage{graphicx}
\usepackage{epsfig}
\usepackage{dcolumn}
\usepackage{bm}
\usepackage{color}
\newcommand {\mev} {\hbox{MeV}/c^2}
\newcommand {\gev} {\hbox{GeV}/c^2}
\newcommand {\A} {A^{0}}

\newcommand {\pip} {\pi^{+}}
\newcommand {\pim} {\pi^{-}}
\newcommand {\pio} {\pi^{0}}
\newcommand {\pippim} {\pi^{+}\pi^{-}}

\newcommand {\mup} {\mu^{+}}
\newcommand {\mum} {\mu^{-}}
\newcommand {\mupmum} {\mu^{+}\mu^{-}}

\newcommand {\jpsi} {J/\psi}
\newcommand {\psip} {\psi^\prime}

\begin{document}

\preprint{\vbox{ \hbox{   }
}}
\title{
Search for a light exotic particle in $J/\psi$ radiative
decays}

\author{
{
M.~Ablikim$^{1}$, M.~N.~Achasov$^{5}$, D.~J.~Ambrose$^{40}$, F.~F.~An$^{1}$, Q.~An$^{41}$, Z.~H.~An$^{1}$, J.~Z.~Bai$^{1}$, R.~B. Ferroli$^{18}$, Y.~Ban$^{27}$, J.~Becker$^{2}$
% ...
\\
\vspace{0.2cm}
(BESIII Collaboration)\\
\vspace{0.2cm} {\it
$^{1}$ Institute of High Energy Physics, Beijing 100049, P. R. China\\
$^{2}$ Bochum Ruhr-University, 44780 Bochum, Germany\\
% ...
$^{5}$ G.I. Budker Institute of Nuclear Physics SB RAS (BINP), Novosibirsk 630090, Russia\\
% ...
$^{18}$ INFN Laboratori Nazionali di Frascati , Frascati, Italy\\
% ...
$^{27}$ Peking University, Beijing 100871, P. R. China\\
% ...
$^{40}$ University of Rochester, Rochester, New York 14627, USA\\
$^{41}$ University of Science and Technology of China, Hefei 230026, P. R. China\\
% ...
}}
\vspace{0.1cm} }

\begin{abstract}
Using a data sample containing $1.06\times 10^8$ $\psip$ events
collected with the BESIII detector at the BEPCII electron-positron
collider, we search for a light exotic particle $X$ in the
process $\psip\rightarrow \pi^{+}\pi^{-}J/\psi$, $J/\psi
\rightarrow \gamma X$, $X \rightarrow \mu^{+}\mu^{-}$. This light particle $X$ could be a Higgs-like boson $A^0$, a spin-1 $U$ boson, or a pseudoscalar sgoldstino particle.
In this analysis, we find no evidence for any
$\mu^+\mu^-$ mass peak between the mass threshold and 3.0 $\hbox{GeV}/c^2$. We
set 90\%-confidence-level upper limits on the product-branching
fractions for $J/\psi \rightarrow \gamma A^{0}$, $A^{0}
\rightarrow \mu^{+}\mu^{-}$ which range from $4\times10^{-7}$ to
$2.1\times10^{-5}$, depending on the mass of $A^0$, for $M(A^0)<
3.0$ $\hbox{GeV}/c^2$. Only one event is seen in the mass region below 255 $\hbox{MeV}/c^2$ and this has a $\mu^+\mu^-$ mass of 213.3 $\hbox{MeV}/c^2$ and the
product branching fraction upper limit $5\times10^{-7}$.
\end{abstract}


\maketitle
\vspace{2cm}

The fundamental nature of mass and dark matter remain among the great mysteries and challenges of science. The Higgs mechanism is a theoretically appealing way to account for masses of elementary particles~\cite{ref1}. A light Higgs-like pseudoscalar boson $\A$ is predicted in the
next-minimal supersymmetric extension of the standard
model~\cite{higgs1,higgs2,higgs3}. A neutral spin-1 boson $U$ in the framework
of the supersymmetric standard model extension is predicted to play an essential role in the annihilations of dark matter~\cite{ref2,ref3,ref4}. Astrophysical observations by PAMELA~\cite{ref5} and ATIC~\cite{ref6} have been interpreted as being due to dark matter annihilation mediated by a light-gauge $U$ boson~\cite{ref7} which couples to standard model particles. The HyperCP
experiment~\cite{barbar10} observed three anomalous
$\Sigma^{+}\rightarrow p \mup\mum$ events with $\mup\mum $
invariant mass clustered around $214.3~\mev$ which are consistent with the process $\Sigma^{+}\rightarrow pX, X\rightarrow\mup\mum$. A particle with these properties could be the pseudoscalar sgoldstino particle~\cite{ref8} in various supersymmetric models~\cite{ref9}, or a light pseudoscalar Higgs-like boson $\A$~\cite{barbar11}, a vector $U$ boson~\cite{ref10} as described above. The lifetime for the pseudoscalar particle case is estimated to be $10^{-14}$~s~\cite{a0time}; which for the $U$ boson depends on the mass of the boson and is smaller than $10^{-14}$~s when the mass of the $U$ boson is more than 100 $\mev$~\cite{utime}.

The D0~\cite{ref11}, CMS~\cite{cms}, LEP~\cite{ref12}, CLEO~\cite{cleo}, BaBar~\cite{barbara0,barbarnewa0} and Belle~\cite{belle} experiments have searched for light-dilepton resonance production using data from $p\bar{p}$ collisions, $e^+e^-$ collisions and b-quark decays. No evidence for a signal of new physics has been found.
It remains important to check the possibility that a particle of these types couples to the $c$-quark and leptons. The branching fraction of $\jpsi\rightarrow \gamma \A$ is expected to be around the $10^{-9}$ to $10^{-7}$ level~\cite{higgs3}. The only search for this kind of particle from charmonium decay was done by the Crystal Ball experiment where from fits to the $\gamma$ recoil energy spectrum, they set branching fraction upper limits of $\jpsi \rightarrow \gamma \A$ which are less than $1.4\times 10^{-5}$(90\% C.L.) for $M(\A)<1.0~\gev$~\cite{cryst}.

The couplings of the Higgs to fermions are proportional to the fermion masses.
For an $\A$ boson with mass below the $\tau$-pair threshold, the
decay $\A\rightarrow \mupmum$ is expected to be dominant.
We use the process $\psip\rightarrow
\pi^{+}\pi^{-}J/\psi$, $J/\psi \rightarrow \gamma A^{0}$, $A^{0}
\rightarrow \mu^{+}\mu^{-}$ to search for an $\A$ with the BESIII
detector~\cite{detector1} at the BEPCII electron-positron
collider~\cite{bepc2}. This $\A$ search is also sensitive to a light spin-1 $U$ boson or a pseudoscalar sgoldstino particle. We assume the $A^0$ particle is a pseudoscalar (or scalar) particle which has narrow width and negligible decay time.


BEPCII is a double-ring $e^{+}e^{-}$ collider with a design peak
luminosity of $10^{33}$~cm$^{-2}s^{-1}$. The BESIII detector is
based on a large $1$-Tesla solenoid magnet and covers $93\%$ of
the total $4\pi$ solid angle surrounding the $e^{+}e^{-}$
collision point with four major detection systems: (1) A
small-cell, helium-based main drift chamber with 43 layers
that provide an average single-hit resolution of 135~$\mu$m,
charged-particle momentum resolution of 0.5\% at 1~GeV/$c$, and a
$dE/dx$ resolution that is better than 6\%. (2) An electromagnetic
calorimeter (EMC) consisting of 6240 CsI(Tl) crystals configured
in a cylindrical structure (barrel) and two end caps. The energy
resolution for 1.0~GeV $\gamma$-rays is 2.5\% in the barrel and
5\% in the end caps, and the position resolution is 6~mm in the
barrel and 9~mm in the end caps. (3) A time-of-flight system
constructed of 5-cm-thick plastic scintillators, with 176 pieces
of 2.4~m long counters arranged in a two layer barrel and 96
fan-shaped counters in the end cap regions. The barrel (end cap)
time resolution of 80~ps (110~ps) provides $2\sigma$ $K/\pi$
separation for momenta up to $\sim 1.0$~GeV$/c$. (4) Muon
identification is provided by 1000~m$^2$ of resistive plate
chambers that are interspersed in the magnet's iron flux
return (MUC). Nine barrel and eight end cap layers provide 2~cm
position resolution for penetrating particles.

The analysis is based on $1.06\times 10^8$ events collected at the
peak of the $\psip$ resonance. The number of $\psip$ events was
determined by counting inclusive hadronic events as described in
Ref.~\cite{npsp} with an estimated uncertainty of 4\%. Monte Carlo
(MC) events are simulated with the {\sc geant4}
program~\cite{geant4} and experimentally determined resolutions of
the wires and counters in the detector.


For the event selection, we first require two positive and two
negative charged tracks and at least one good photon. We also use
$\mu$ identification information, veto $\pio$s, place restrictions on the mass recoiling from the $\pip\pim$ system, and apply kinematic
constraints. The dominant backgrounds are from $\psip\rightarrow
\pi^{+}\pi^{-}J/\psi$, with $J/\psi \rightarrow \gamma \pippim$,  $J/\psi \rightarrow \rho\pi \rightarrow \pip\pim\pio$ or $J/\psi \rightarrow l^{+}l^{-}$. The kinematic constraints are especially effective for removing backgrounds from $J/\psi \rightarrow \pip\pim\pio$ and $J/\psi \rightarrow l^{+}l^{-}$ decays.
A $\pio$ veto is used to reject $J/\psi \rightarrow \pip\pim\pio$ events.
The selection requirements used for the $\pio$
veto, the $\pi^{+}\pi^{-}$ recoil mass requirement and the kinematic fit quality are optimized
for the assumption that the branching fraction for $J/\psi \rightarrow \gamma A^{0}$,
$A^{0} \rightarrow \mu^{+}\mu^{-}$ is at the $10^{-6}$ level and using $s/\sqrt{(s+b)}$ as a figure of merit,
where $s$ is the expected number of signal events and $b$ is the number of background events.
The track selection criteria are standard in BESIII analysis.

Candidate photons are energy clusters in the EMC which: (1) are
within the fiducial region of the EMC ($|\cos\theta_{\gamma}| <
0.8$ for the barrel and $0.84 <|\cos\theta_{\gamma}|< 0.92$ for
the end caps); (2) are more than 20 degrees away from the
extrapolated position of the closest charged track; (3) have a
pulse time that is consistent with being produced together with
the charged-track candidates.

Charged track candidates are required to originate from the
interaction point, $V_{xy}$ =$\sqrt{V_{x}^2 + V_{y}^2} <
1~\hbox{cm}$, $|V_{z}| < 10~\hbox{cm}$, where $V_{x}$, $V_{y}$,
and $V_{z}$ are the $x$, $y$, and $z$ coordinates of the point of
closest approach to the interaction point. The tracks are also
required to be within the polar angle region $|\cos\theta|<0.93$.

Candidate muons are charged tracks in the active area of the
barrel MUC ($|\cos\theta|<0.75$) with: momentum higher than
0.7~GeV/$c$; energy deposition in the EMC between 0.15~GeV and
0.26~GeV; $E/p$ (EMC energy over main drift chamber momentum) less than 0.5; and at
least three associated hit layers in the MUC. For tracks in the
momentum range $0.8~\hbox{GeV}/c <p< 1.15~\hbox{GeV}/c$, the MUC
penetration depth is required to be greater than $(70p-40)$~cm
($p$ in GeV/$c$); for tracks with $p>1.15~\hbox{GeV}/c$, the
penetration depth is required to be more than 41~cm. Tracks with
momentum above $0.8~\hbox{GeV}/c$ are removed if the fit to
the MUC hits either fails or gives a poor fit result. The
muon identification(PID) single-track efficiency is typically 65\%, and the $\pi$
fake rate is less than 5\%/track.
%

If there are multiple photons with energy above 25~MeV, we
reject the event if any pair of these photons has an invariant
mass within 40~MeV/$c^2$ of $m_{\pi^0}$. For the multiphoton
events that remain, the $\gamma$ with the highest energy is
selected as the photon used in the analysis. The pair of
oppositely charged tracks with recoil mass closest to the $J/\psi$
mass is assigned as the $\pi^{+}$ and $\pi^{-}$ and the other two
tracks as the $\mu^{+}$ and $\mu^{-}$. At least one of the tracks
assigned as a muon is required to satisfy the $\mu$-PID
criteria. We select events with a $\pi^{+}\pi^{-}$ recoil mass in
the range between 3.092~$\gev$ and 3.102~$\gev$ and perform a
four-constraint energy-momentum conserving kinematic fit using the
selected $\gamma$ and four charged tracks. We require
$\chi^{2}<40$ and $M(\mupmum)<3.02~\gev$.

\begin{figure*}[htbp]
\begin{center}
\includegraphics[width=8.5cm]{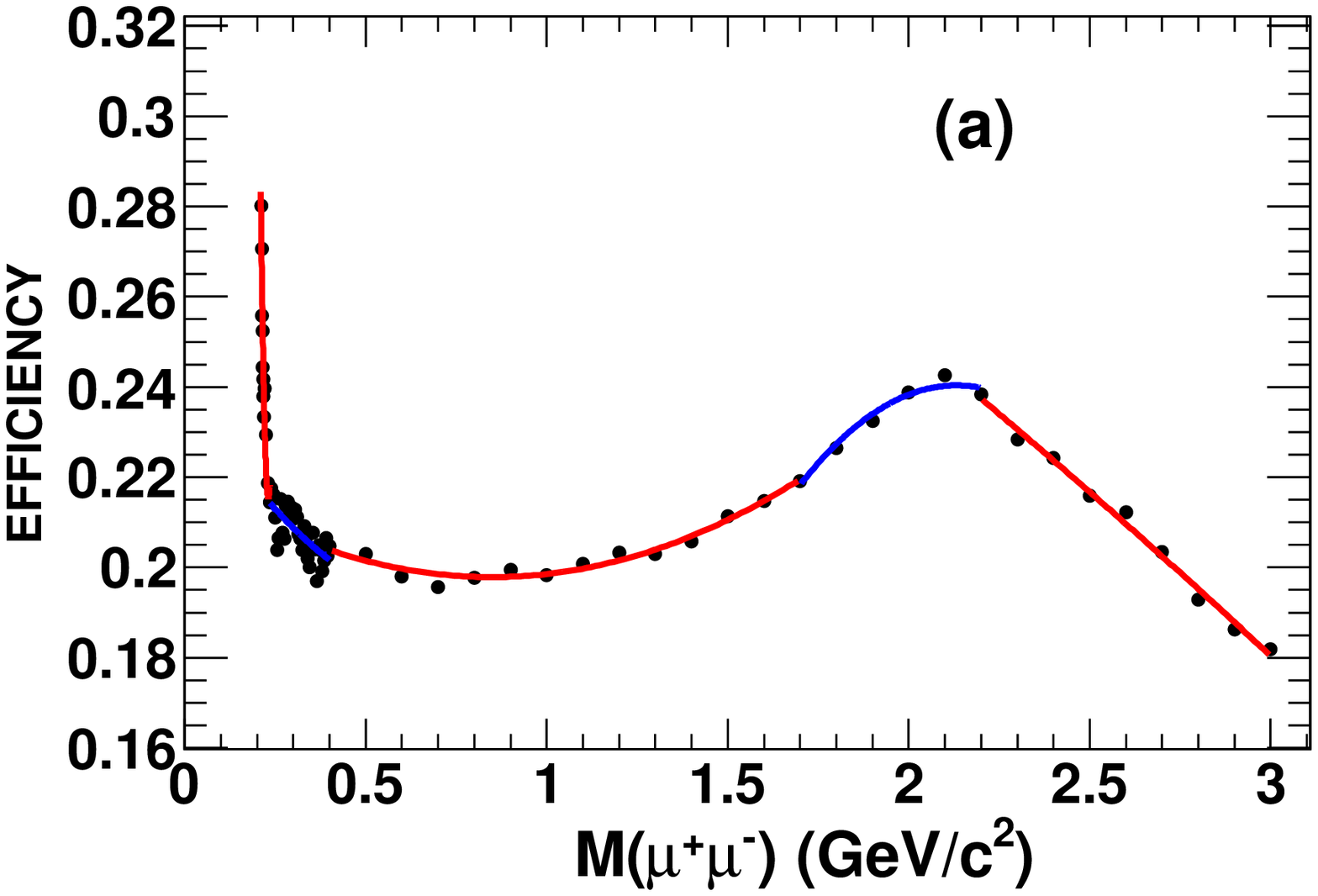}
 \put(-180,130){}
\includegraphics[width=8.5cm]{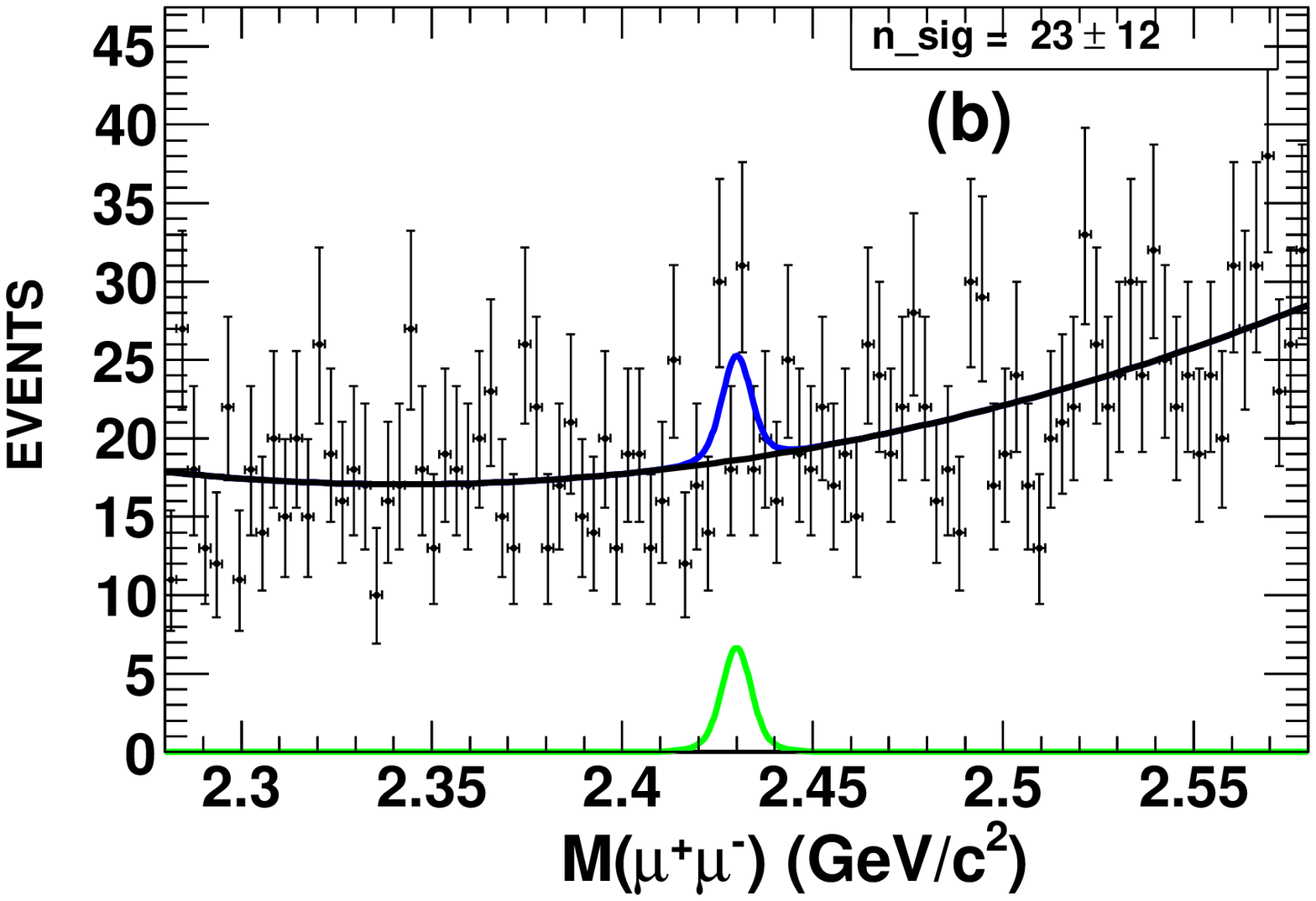}
 \put(-100,130){}
\end{center}
\caption{(a) The event selection efficiency for $\psip\rightarrow
\pi^{+}\pi^{-}J/\psi$, $J/\psi \rightarrow \gamma A^{0}$, $A^{0}
\rightarrow \mu^{+}\mu^{-}$. (b) The fit to the invariant mass spectrum $M(\mupmum)$ in
the 5 $\mev$ wide interval centered at 2.43 $\gev$ showing the total fit
result and the background-subtracted signal.}
\label{fitone}
\end{figure*}

Simulations
where the $\A$ width is set to zero and the mass is set at 71
different values which range from 0.212~$\gev$ to 3.0~$\gev$
indicate that the selection efficiency varies between 28\% and
18\%, depending on the mass of $\A$, as shown in Fig.~\ref{fitone}(a).
The simulation is done for 1~$\mev$ $\A$ mass steps for $M(\A)$ between
0.212~$\gev$ to 0.22~$\gev$, 5~$\mev$ steps for $M(\A)$ between 0.22~$\gev$
to 0.4~$\gev$ and 100~$\mev$ steps for $M(\A)$ above 0.4~$\gev$.
We fit the resulting efficiency values piecewise with second-order polynomial shapes to get the $\A$-mass-dependent
efficiency which includes any bias caused by the fit. The $\A$ mass resolution
determined from the MC simulation increases with $\A$ mass, ranging
from 0.1 $\mev$ near the low-mass threshold to about 5 $\mev$ for masses
near 3.0 $\gev$. The efficiencies for spin-1 $U$ production are the same as
those for the $A^0$ to within a few percent.

%

\begin{figure*}[]
\begin{center}
 \includegraphics[width=1.0\textwidth]{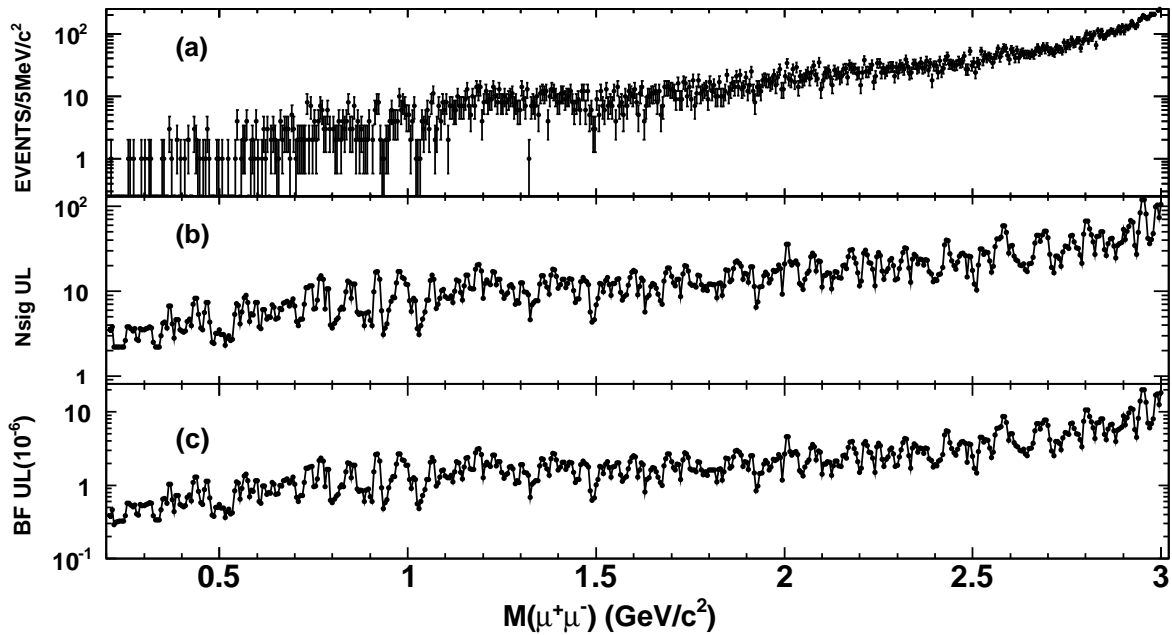}
\end{center}
\caption{(a) The $\mu^+\mu^-$ invariant mass spectrum for the
selected $\psip\rightarrow \pi^{+}\pi^{-}J/\psi$, $J/\psi\to
\gamma\mupmum$ events; (b) 90\% C.L. upper
limits on the number of signal events (Nsig UL) as a function of
the $\mup\mum$ invariant mass; (c) upper limits on the branching
fractions (BF UL) for $\jpsi \rightarrow \gamma A^{0}$, $A^{0}
\rightarrow \mu^{+}\mu^{-}$ at the 90\% C.L.} \label{data}
\end{figure*}

The $\mup\mum$ mass distribution of selected data events is shown
in Fig.~\ref{data}(a). Over the entire mass range, from threshold
to 3.0~$\gev$, there is no evident narrow peak. Below 255~$\mev$,
there is only one event, with $\mupmum$ invariant mass of
213.3~$\mev$. The expected $\A$ mass resolution is about 0.2
$\mev$ for $M(A^0)=213.3~\mev$ and the major background in
this region comes from $\psip\rightarrow \pi^{+}\pi^{-}J/\psi$,
$J/\psi \rightarrow \gamma \pi^{+}\pi^{-}$. The expected number of
background events in the mass region near $213.3~\mev$ is about
0.2/$\mev$; the observation of one event in this region is
consistent with that at background level.

To set upper limits on the production rates for different masses,
we do unbinned maximum-likelihood fits to $\sim 300~\mev$-wide ranges of the $\mup\mum$ invariant
mass spectrum where the mass of the $\A$ peak is restricted to be
within a series of 5 $\mev$-wide intervals near the center of the range.
In each fit, we use a MC-determined shape for the $\A$ signal, and
for the background shape, we use a polynomial.
We do not find any significant signal
and set Bayesian upper limits on the signal yield in each 5~MeV/$c^2$ interval.
Figure.~\ref{fitone}(b) shows a typical fit to the $\mupmum$ invariant mass spectrum in
the 5 $\mev$-wide interval centered at 2.43 $\gev$.


We use different fit ranges, polynomial background shapes of
different orders, and MC signal shapes for different $\A$
mass values to estimate the fit-related
systematic error on the signal yield in each mass interval.
We first fit using the MC signal shape from the nearest
generated $\A$ mass for the signal shape with a second-order
polynomial to represent the background
shape. We then increase and decrease the edges of fit range by
$\pm 5~\mev$, use signal shapes from the MC fits that
are one step lower and one step higher than the nearest one, and use
first- and third-order polynomial shapes for the background.
Each fit is required to converge.
For each mass interval, the fit variation that produces
the largest number of signal events is used to determine
the 90\% C.L. upper limits, which are shown in Fig.~\ref{data}(b).

The systematic errors in the $J/\psi\to \gamma \A$, $\A\to
\mup\mum$ product-branching-fraction measurement are summarized in
Table~\ref{allerr}; these include contributions from tracking,
particle identification, photon selection, kinematic fit,
$\pip\pim$ recoil mass requirement, and $\pi^0$ veto. The
uncertainty of the number of $\psi^\prime$ events is
4\%~\cite{npsp} and that of the $\psi^\prime\to \pip\pim J/\psi$
branching fraction is 1.2\%~\cite{pdg10}.

\begin {table}[htbp]
\caption {The individual contributions to the total relative
systematic error (\%) in the product branching fraction
measurement.} \label{total error}
\begin {center}
\begin {tabular}{cc}
\hline
 Source                     & ~~~~~Error~~~~~ \\ \hline
Tracking efficiency        & 4.0   \\
Particle identification    & 2.5   \\
Kinematic fit              & 3.2   \\
$\gamma$ efficiency        & 1.0   \\
$\pi^{+}\pi^{-}$ recoil mass & 1.2   \\
$\pio$ veto                     & 2.0   \\
Number of $\psi^\prime$s        & 4.0   \\
${\mathcal B}(\psi^\prime\to \pip\pim J/\psi)$ & 1.2  \\ \hline
 Total                          & 7.5   \\\hline
\end {tabular}
\label{allerr}
\end {center}
\end {table}

The uncertainty due to data-MC differences in the charged-tracking
efficiency is 1\% per track and added linearly. This is determined
from high-statistics, low-background samples of $J/\psi\to \rho\pi$ and $J/\psi\to
p\bar{p}\pi^{+}\pi^{-}$ events. In this analysis, there are four
charged tracks, and the relative systematic error is 4\%.

The uncertainty
due to the photon reconstruction is determined to be 1\% for
each photon using three different methods as described in Ref.~\cite{xgf}.
These include a missing photon and $\pio$ decay angle method
using  a clean sample of $\psip\rightarrow \pi^{+}\pi^{-}\jpsi, \jpsi\rightarrow\rho^0 \pio$
events, and a missing $\pio$ method using
$\psip\rightarrow\pio\pio\jpsi, \jpsi\rightarrow l^+l^-$ events.

The uncertainties due to muon identification are determined
from studies of a sample of radiative muon pair events that
contain one photon. We determine muon PID probabilities for 0.3 GeV/$c$
steps in track momentum and determine that the efficiency is about
65\% per track; the data-MC differences in efficiency are
less than 4\% per track. Since we require only one
muon to satisfy the identification criteria, the PID-related
systematic error is less than 2.5\%.

The systematic uncertainty associated with the kinematic fit is
determined by applying a similar kinematic fit to MC and data
samples of $\psip\rightarrow \pip\pim\jpsi$, $\jpsi\rightarrow
\pip\pim\pio$, $\pio\rightarrow \gamma\gamma$ events. In the event
selection for this study, if there are more than two candidate
$\gamma$s we use the most energetic $\gamma$ together with the one
that has the best one-constraint fit to $\pio\rightarrow \gamma\gamma$. From
data-MC differences for these events, the systematic error
associated with the kinematic fit is determined to be 3.2\%.

It is unlikely for  signal events to have an $M(\gamma\gamma)$ value that is
near $m_\pio$; the efficiency reduction caused by the $\pio$ veto
is less than 3\%.
The systematic error associated with the $\pio$ veto is studied
with samples of $\psip\rightarrow \gamma \chi_{cJ} \rightarrow
\gamma\phi\phi \rightarrow \gamma 2(K^{+}K^{-})$ and
$\psip\rightarrow \pi^{+}\pi^{-}J/\psi$, $J/\psi\rightarrow \gamma
f_2(1270)$, $f_2(1270)\rightarrow \pi^{+}\pi^{-}$ events from both
MC simulation and data. From $\psip\rightarrow \gamma \chi_{cJ}
\rightarrow \gamma\phi\phi$, we determine the effect of the $\pio$
veto cut on the efficiency. For the second sample, we fit the
$\pip\pim$ mass spectrum to get the number of $f_2(1270)
\rightarrow \pi^+\pi^-$ events with and without the $\pio$ veto
cut. Data and MC efficiency differences for the $\pio$ veto are
found to be less than 1.7\% from the first channel and less than
2.0\% from the second channel. We use 2\% as the systematic error
due to the $M(\gamma\gamma$) requirement.

The systematic error caused by the $\pi^{+}\pi^{-}$ recoil mass
requirement is analyzed with the sample of $\psip\rightarrow
\pi^{+}\pi^{-}J/\psi$, $J/\psi\rightarrow \mu^{+}\mu^{-}$ events in the data
and from MC simulation. From the numbers of events with and without the
recoil mass requirement we determine data and MC efficiency
difference to be less than 1.2\%.

The systematic errors are summarized in Table~\ref{allerr}, each of these are the largest errors over the entire
$M(\mupmum)$ range. Assuming the errors from all sources are independent, the total
error is determined from the quadrature sum to be 7.5$\%$.


We determine the upper limit on the branching fractions of $\jpsi
\rightarrow \gamma A^{0}$, $A^{0} \rightarrow \mu^{+}\mu^{-}$ from
the relation

\begin{equation}
 {\mathcal B}<\frac{\rm Nsig(UL)/\varepsilon}
 { N(\psip)\times {\mathcal B}(\psip\to \pip\pim\jpsi)
 \times (1-\sigma)},
 \label{equlimit}
 \end{equation}
where $\rm Nsig(UL)$, shown in Fig.~\ref{data}(b), is the upper limit on the number of signal events in each $M(\mupmum)$ bin after consideration of the mass
fitting systematic errors;
$\varepsilon$ is the $\A$-mass-dependent selection efficiency
determined from MC simulation; $N(\psip) = 1.06\times 10^{8}$ is
the number of $\psip$ events~\cite{npsp} and ${\mathcal B}(\psip
\to \pip\pim\jpsi)=(33.6\pm 0.4)\%$ is the PDG world
average~\cite{pdg10}. The upper limit is increased by a factor of
$1/(1-\sigma)$, where $\sigma$ is the total systematic error
($7.5\%$) to give a conservative result. The resulting ${\mathcal
B}(J/\psi \rightarrow \gamma A^{0}) \times {\mathcal B}(A^{0}
\rightarrow \mu^{+}\mu^{-})$ upper-limit values range from $4\times10^{-7}$ for an $\A$
mass near threshold to $2.1\times 10^{-5}$ for $M(\A)$ near 3.0~$\gev$,
and is shown in Fig.~\ref{data}(c). The branching fraction upper
limit is less than $10^{-6}$ for all $M(\A)$ values below
0.36~$\gev$, and is less than $10^{-5}$ for all masses below
2.79~$\gev$.


In summary, we have searched for a light exotic particle at
BESIII. No evidence is observed and upper limits on the product
branching fractions for $J/\psi \rightarrow \gamma A^{0}$, $A^{0}
\rightarrow \mu^{+}\mu^{-}$ range from $4 \times 10^{-7}$ to
$2.1\times 10^{-5}$, depending on the mass of the $\A$, are
established. These limits are new stringent
experimental results from charmonium decays, and can rule out much of the parameter space in theoretical models~\cite{fayet09}. Only one event is observed in the low-mass
region below 255~$\mev$, with a $\mupmum$ mass of 213.3~$\mev$.
For $M(\A)<255~\mev$, including the 214.3~$\mev$ mass value of the
anomalous HyperCP $\Sigma^{+}\rightarrow p \mup\mum$ events, the
product branching fraction upper limit is $5 \times 10^{-7}$ at
the 90\% C.L. Although these branching fraction upper limits are computed for a spin-0 particle, they are the same, to within a few percent, for a spin-1 particle.

\acknowledgments

The BESIII Collaboration thanks the staff of BEPCII and the computing center for their hard efforts. This work is supported in part by the Ministry of Science and Technology of China under Contract No. 2009CB825200; National Natural Science Foundation of China (NSFC) under Contracts Nos. 10625524, 10821063, 10825524, 10835001, 10935007; Joint Funds of the National Natural Science Foundation of China under Contract No. 11079008; the Chinese Academy of Sciences (CAS) Large-Scale Scientific Facility Program; CAS under Contract Nos. KJCX2-YW-N29, KJCX2-YW-N45; 100 Talents Program of CAS; Istituto Nazionale di Fisica Nucleare, Italy; U. S. Department of Energy under Contracts Nos. DE-FG02-04ER41291, DE-FG02-91ER40682, DE-FG02-94ER40823; U.S. National Science Foundation; University of Groningen (RuG) and the Helmholtzzentrum fuer Schwerionenforschung GmbH (GSI), Darmstadt; WCU Program of National Research Foundation of Korea under Contract No. R32-2008-000-10155-0.


\begin{thebibliography}{99}

\bibitem{ref1}
P.~W.~Higgs, Phys.\ Rev.\ Lett.\ {\bf 13}, 508 (1964).

\bibitem{higgs1}
R.~Dermisek and J.~F.~Gunion,
Phys.\ Rev.\ Lett.\ {\bf 95}, 041801 (2005).

\bibitem{higgs2}
R.~Dermisek and J.~F.~Gunion,
Phys.\ Rev.\ D {\bf 77}, 015013 (2008).

\bibitem{higgs3}
R.~Dermisek, J.~F.~Gunion and B.~McElrath,
Phys.\ Rev.\ D {\bf 76}, 051105 (2007).

\bibitem{ref2}
P.~Fayet, Phys.\ Lett.\ B {\bf 95}, 285 (1980).

\bibitem{ref3}
C.~Boehm and P.~Fayet, Nucl.\ Phys.\ B {\bf 683}, 219 (2004).

\bibitem{ref4}
P.~Fayet,
Phys.\ Rev.\ D {\bf 74}, 054034 (2006).

\bibitem{ref5}
O.~Adriani {\it et al.} [PAMELA Collaboration], Nature {\bf 458},
607 (2009).

\bibitem{ref6}
J.~Chang {\it et al.} [ATIC Collaboration], Nature {\bf 456}, 362
(2008).

\bibitem{ref7}
M.~Pospelov, A.~Ritz and M.~B.~Voloshin, Phys.\ Lett.\ B
{\bf 662}, 53 (2008); N.~Arkani-Hamed and N.~Weiner, JHEP
{\bf 0812}£¬104 (2008).

\bibitem{barbar10}
H.~K.~Park {\it et al.}  [HyperCP Collaboration],
Phys.\ Rev.\ Lett.\  {\bf 94}, 021801 (2005).

\bibitem{ref8}
D.~S.~Gorbunov and V.~A.~Rubakov, Phys.\ Rev.\ D {\bf 73},
035002 (2006).

\bibitem{ref9}
J.~Ellis, K.~Enqvist and D.~Nanopoulos, Phys.\ Lett.\ B
{\bf 147}, 99 (1984); T.~Bhattacharya and P.~Roy, Phys.\ Rev.\
D {\bf 38}, 2284 (1988); G.~Giudice and R.~Rattazzi, Phys.\
Rep.\ {\bf 322}, 419 (1999).

\bibitem{barbar11}
X.~G.~He, J.~Tandean and G.~Valencia,
Phys.\ Rev.\ Lett.\  {\bf 98}, 081802 (2007).


\bibitem{ref10}
M.~Reece and L.~T.~Wang, JHEP {\bf 0907}, 51 (2009);
M.~Pospelov, Phys.\ Rev.\ D {\bf 80}, 095002 (2009); C.~H.~Chen,
C.~Q.~Geng and C.~W.~Kao, Phys.\ Lett.\ B {\bf 663}, 400
(2008).

\bibitem{a0time}
C.~Q.~Geng and Y.~K.~Hsiao,
Phys.\ Lett.\ B {\bf 632}, 215 (2006).

\bibitem{utime}
P.~Fayet,
Nucl.\ Phys.\ B {\bf 187}, 184 (1981).

\bibitem{ref11}
V.~M.~Abazov {\it et al.} [D0 Collaboration], Phys.\ Rev.\ Lett.\
{\bf 103}, 061801 (2009).

\bibitem{cms}
S.~Chatrchyan {\it et al.} [CMS Collaboration],
JHEP {\bf 1107}, 098, (2011).

\bibitem{ref12}
S. Schael {\it et al.} [ALEPH Collaboration], JHEP {\bf 1005}, 049 (2010).


\bibitem{cleo} W.~Love {\it et al.} [CLEO Collaboration],
Phys.\ Rev.\ Lett.\ {\bf 101}, 151802 (2008).

\bibitem{barbara0}
B.~Aubert {\it et al.} [BABAR Collaboration],
Phys.\ Rev.\ Lett.\ {\bf 103}, 081803 (2009).

\bibitem{barbarnewa0}
B.~Aubert {\it et al.}  [BABAR Collaboration],
Phys.\ Rev.\ Lett.\  {\bf 103}, 181801 (2009).

\bibitem{belle}
H.~J.~Hyun {\it et al.} [Belle Collaboration],
Phys.\ Rev.\ Lett.\ {\bf 105}, 091801 (2010).

\bibitem{cryst}
C.~Edwards {\it et al.}, [Crystal Ball Collaboration,], Phys.\ Rev.\ Lett.\ {\bf 48}, 903 (1982).

\bibitem{detector1}
M.~Ablikim {\it et al.} [BESIII Collaboration],
Nucl.\ Instrum.\ Meth.\  A {\bf 614}, 345 (2010).

\bibitem{bepc2}
J.~Z.~Bai {\it et al.} [BES Collaboration], Nucl.\ Instrum.\ Meth.\ A {\bf 344}, 319 (1994); Nucl.\ Instrum.\ Meth.\ A {\bf 458}, 627 (2001).

\bibitem{npsp} M.~Ablikim {\it et al.} [BESIII Collaboration],
Phys.\ Rev.\ D {\bf 81}, 052005 (2010).

\bibitem{geant4}
S.~Agostinelli {\it et al.} [GEANT4 Collaboration],
Nucl.\ Instrum.\ Meth.\ A {\bf 506}, 250 (2003).

\bibitem{pdg10}
K.~Nakamura {\it et al.}  [Particle Data Group],
J.\ Phys.\ G {\bf 37}, 075021 (2010).

\bibitem{xgf}
M.~Ablikim {\it et al.} [BESIII Collaboration],
Phys.\ Rev.\ D {\bf 83}, 112005 (2011).

\bibitem{fayet09}
P.~Fayet,
Phys.\ Lett.\ B {\bf 675}, 267 (2009).

\end{thebibliography}
\end{document}